\documentclass[prl,twocolumn,showpacs,superscriptaddress]{revtex4-1}
\usepackage{graphicx}
\usepackage{epsfig}
\usepackage{epsf}
\usepackage{amssymb}
\usepackage{amsmath}
\usepackage{amsthm}
\usepackage{multirow}
\usepackage[dvipdfm]{hyperref}

\newcommand{\bra}[1]{\langle #1|}
\newcommand{\ket}[1]{|#1\rangle}

\newcommand{\be}{\begin{equation}}
\newcommand{\ee}{\end{equation}}
\newcommand{\bea}{\begin{eqnarray}}
\newcommand{\eea}{\end{eqnarray}}

\usepackage[usenames]{color}
\definecolor{Red}{rgb}{1,0,0}
\definecolor{Blue}{rgb}{0,0,1}

\begin{document}

% Include your paper's title here
\title{Experimental Estimation of Average Fidelity of a Clifford Gate on a 7-qubit Quantum Processor}
\author{Dawei Lu}
\affiliation{Institute for Quantum Computing and Department of Physics,
University of Waterloo, Waterloo, Ontario N2L 3G1, Canada}

\author{Hang Li}
\affiliation{Institute for Quantum Computing and Department of Physics,
University of Waterloo, Waterloo, Ontario N2L 3G1, Canada}
\affiliation{State Key Laboratory of Low-Dimensional Quantum Physics and Department of Physics, Tsinghua University, Beijing 100084, China}
\affiliation{Collaborative Innovation Center of Quantum Matter, Beijing 100084, China}

\author{Denis-Alexandre Trottier}
\affiliation{Institute for Quantum Computing and Department of Physics,
University of Waterloo, Waterloo, Ontario N2L 3G1, Canada}

\author{Jun Li}
\affiliation{Institute for Quantum Computing and Department of Physics,
University of Waterloo, Waterloo, Ontario N2L 3G1, Canada}
\affiliation{Department of Modern Physics, University of Science
and Technology of China, Hefei, Anhui 230026, China}

\author{Aharon Brodutch}
\affiliation{Institute for Quantum Computing and Department of Physics,
University of Waterloo, Waterloo, Ontario N2L 3G1, Canada}

\author{Anthony P. Krismanich}
\author{Ahmad Ghavami}
\author{Gary I. Dmitrienko}
\affiliation{Department of Chemistry, University of Waterloo, Waterloo,
Ontario N2L 3G1, Canada}
\author{Guilu Long}
\affiliation{State Key Laboratory of Low-Dimensional Quantum Physics and Department of Physics, Tsinghua University, Beijing 100084, China}
\affiliation{Collaborative Innovation Center of Quantum Matter, Beijing 100084, China}
\author{Jonathan Baugh}
\affiliation{Institute for Quantum Computing and Department of Physics,
University of Waterloo, Waterloo, Ontario N2L 3G1, Canada}
\affiliation{Department of Chemistry, University of Waterloo, Waterloo,
Ontario N2L 3G1, Canada}
\author{Raymond Laflamme}
\email{laflamme@iqc.ca}
\affiliation{Institute for Quantum Computing and Department of Physics,
University of Waterloo, Waterloo, Ontario N2L 3G1, Canada}
\affiliation{Perimeter Institute for Theoretical Physics, Waterloo, Ontario N2L 2Y5, Canada}
\affiliation{Canadian Institute for Advanced Research, Toronto, Ontario M5G 1Z8, Canada}

\begin{abstract}
Quantum gates in experiment are inherently prone to  errors that need to be characterized before they can be corrected. Full characterization via quantum process tomography is impractical and often unnecessary.
% \B{(Is it better to say impractical here? 'Unnecessary' is only for some particular cases and thus QPT can be replaced  by twirling)}
For most practical purposes,  it is enough to   estimate more  general quantities  such as the average fidelity. Here we use a unitary 2-design and twirling protocol for  efficiently  estimating  the  average fidelity of Clifford gates, to certify a 7-qubit  entangling gate in a nuclear magnetic resonance quantum processor.
%The traditional approach of characterizing a quantum gate via quantum process tomography requires an exponential number of experiments, which is impractical even for moderately large systems. In contrast, the twirling protocol is demonstrated to be able to tackle these problems efficiently, such as estimating the average fidelity of quantum memories and Clifford gates. Here, we certify an essential but complicated Clifford gate in a 7-qubit nuclear magnetic resonance quantum processor using the twirling protocol and sampling method.
Compared with more than $10^8$ experiments required by full process tomography, we conducted 1656 experiments to satisfy a statistical confidence level of 99\%. The average fidelity of this Clifford gate in experiment is 55.1\%, and rises to 87.5\%  if the infidelity due to decoherence is removed. %The experiment spectrum related to this Clifford gate is also in great agreement with the simulated one.
%We state that reliable coherent controls in this system have been achieved in spite of the unavoidable decoherence.
The entire protocol of certifying Clifford gates is efficient and scalable, and can  easily be extended to any general quantum information processor with minor modifications. .
\end{abstract}

\maketitle

\paragraph*{Introduction.}

Benchmarking protocols for characterizing the level of coherent control are  fundamental  in evaluating potential quantum information processing (QIP) devices. They  provide an objective  comparison of quantum control capabilities between diverse QIP devices, and also indicate the prospects of a  given platform  with respect to  fault-tolerant quantum computation \cite{Preskill1998}. The traditional approach of  using  quantum process tomography (QPT) \cite{Chuang1997,Poyatos1997} is useful for completely characterizing  a quantum channel, and has been applied to at most 3-qubit systems in experiment \cite{Childs2001,Weinstein2004,Brien2004,Riebe2006,Chow2009,Bialczak2010,Kim2014,Feng2013}. However, QPT  requires number of measurements that scale exponentially with  number of qubits $n$ ($\approx2^{4n}$), making it impractical  even in relatively small systems. Moreover,for many practical purposes, such as benchmarking, the full description of a particular quantum channel  is not necessary and more accessible properties  of the gates are sufficient. To benchmark a gate it is enough to estimate the distance between the implemented channel and the ideal gate.   Several methods such as randomized benchmarking \cite{Emerson2005,Knill2008,Ryan2009}, twirling \cite{Emerson2007,Dankert2009,Moussa2012}, and Monte Carlo estimations \cite{Flammia2011,Silva2011} have been proposed to evaluate a particular quantum channel in an efficient manner, each with its  own restrictions and drawbacks. Here, in order to benchmark our coherent controls on a 7-qubit nuclear magnetic resonance (NMR) system, we adopted the twirling protocol \cite{Moussa2012} to estimate the average fidelity of an important Clifford gate in QIP. The gate of  interest  generates   maximal coherence from single coherence with the aid of  local rotations, and is  of critical importance to many QIP tasks such as the creation of a cat state in multi-qubit systems. The estimation method is scalable and independent of the number of qubits, and is straightforward to implement in other quantum information processing architectures.

For the twirling protocol we conducted only 1656 experiments compared with about $2.7\times 10^8$ experiments required for fully characterizing the 7-qubit gate via QPT.  The average fidelity of the certified gate is 55.1\% before accounting for decoherence and rises to 87.5\% by separating the decoherence effect out. Moreover, the NMR spectra based on the application of this Clifford gate are in excellent agreement with the simulation results.

\paragraph*{Theory.}
%\R{[I re-wrote the whole 2-desing and twirling part]}

Let  $\mathcal{U}$ be a superoperator representation of  the  Clifford gate $U$ that we want to implement and $\tilde{\mathcal{U}} =\Lambda\circ \mathcal{U}$ be the superoperator representation of the real evolution in the laboratory experiment.  We call $\Lambda$ the noise superoperator and our task is to estimate its  average fidelity with respect to the identity. The method described below is based on twirling \cite{Bennett1996} and the  construction of  a unitary 2-design \cite{Dankert2009}.

Given  a fiducial  pure state  $\ket\psi$, the average fidelity (with respect to the identity) is the quantum fidelity $\bra\psi\Lambda(\ket\psi\bra\psi)\ket\psi$ averaged over all pure states $V\ket\psi$ where $V$ is an arbitrary unitary transformation. Averaging over the entire Hilbert space can be done using the Haar measure   $d\mu(\mathcal{V})$ \cite{Emerson2005},  so
\begin{align} \label{average_Harr}
\bar{F}(\Lambda) = \int d\mu(\mathcal{V}_U) \bra{\psi} \mathcal{V_U}^{\dagger} \circ \Lambda \circ  \mathcal{V}_U(\ket{\psi} \bra{\psi}) \ket{\psi}.
\end{align}
Here $\mathcal{V}$ is the superoperator representation of a unitary $V$ and $\mathcal{V}_U=\mathcal{U}\circ\mathcal{V}$. In this notation it is easy to see that the average fidelity depends only on $\Lambda$ .

Using a unitary 2-design based on the Clifford group, it is possible to simplify Eq. \eqref{average_Harr} to
\begin{align} \label{average_Clifford}
\bar{F}(\Lambda) =  \frac{1}{|\mathcal{C}_n|}\sum_{\mathcal{C}_i\in \mathcal{C}_n}\bra{\psi} \mathcal{C}_i^{\dagger} \circ \Lambda \circ \mathcal{C}_i(\ket{\psi} \bra{\psi}) \ket{\psi},
\end{align}
where $\mathcal{C}_n$  is the $n$-qubit Clifford group $\mathcal{C}_n$.  The average fidelity is therefore equivalent to the fidelity of the average channel
 \begin{align} \label{Cn_twirl}
\bar{\Lambda}_{\mathcal{C}_n} = \frac{1}{|\mathcal{C}_n|}\sum_{\mathcal{C}_i\in \mathcal{C}_1} \mathcal{C}_i^{\dagger} \circ \Lambda \circ \mathcal{C}_i.
\end{align}
This is a depolarizing channel $\bar{\Lambda}_{\mathcal{C}_n}(\rho)=P_0\rho+[1-P_0]\mathcal{I}$ with $P_0$ the  probability for no error. The average fidelity of Eq. \eqref{average_Clifford} is therefore a function of the parameter $P_0$.

To estimate $P_0$ in a scalable  way  we can make use of an identification involving  the $C_1^{\otimes n} \Pi$-twirled channel. This is the channel $\Lambda$ twirled over the composition of the $n$-fold tensor product of the 1-qubit Clifford group $C_1^{\otimes n}$ and  the  permutation group $ \Pi$
  \begin{align} \label{C1_twirl}
\bar{\Lambda}_{\mathcal{C}_1^{\otimes n}\Pi} = \frac{1}{|\mathcal{C}_1^{\otimes n}\Pi|}\sum_{\mathcal{C}_i\in \mathcal{C}_1^{\otimes n}\Pi} \mathcal{C}_i^{\dagger} \circ \Lambda \circ \mathcal{C}_i.
\end{align}
It has a Pauli form
\begin{align} \label{C1_twirlrho}
\bar{\Lambda}_{\mathcal{C}_1^{\otimes n}\Pi}(\rho) = \sum_{w=0}^n \text{Pr}(w) \left ( \frac{1}{3^w \binom{n}{w}} \sum_{i=1}^{3^w \binom{n}{w}} \mathcal{P}_{i,w} \rho \mathcal{P}_{i,w} \right ),
\end{align}
where $\text{Pr}(w)$ is the probability that a Pauli error of weight $w$ occurs.  The identification $P_0=\text{Pr}(0)$  \cite{Silva2008} gives

\begin{align} \label{fidelity_pr}
\bar{F}(\Lambda) = \frac{2^n \text{Pr}(0) +1}{2^n +1}.
\end{align}

The task of finding the average fidelity of the noisy channel $\Lambda$ is now reduced to finding Pr(0), i.e. the probability that the twirled  channel $\bar{\Lambda}_{\mathcal{C}_1^{\otimes n} \Pi}$ does not cause an error.

\begin{figure*}[htb]
\begin{center}
\includegraphics[width=2\columnwidth]{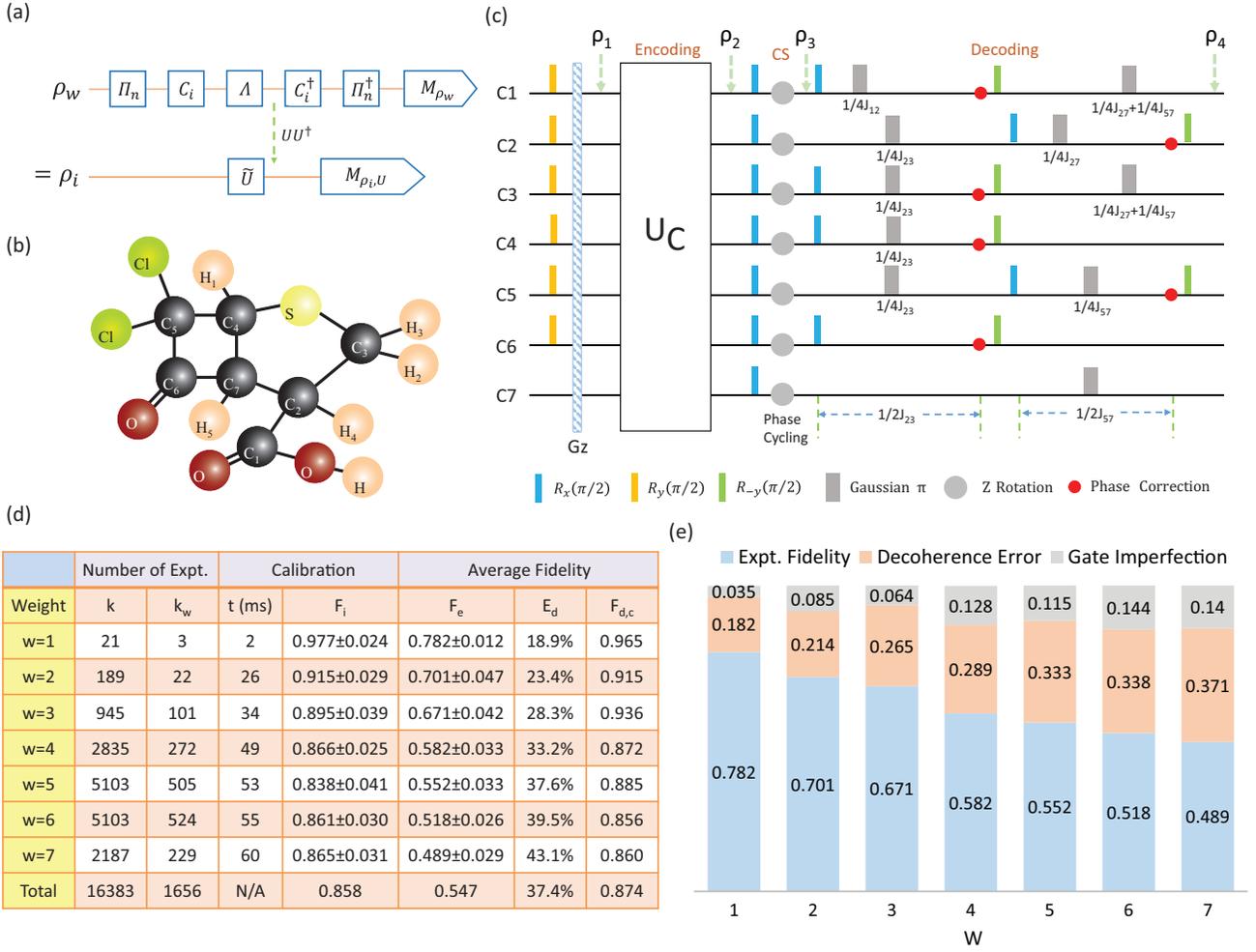}
\end{center}
\setlength{\abovecaptionskip}{-0.35cm}
\caption{\footnotesize{(color online). (a) Twirling protocols for quantum memories (top) and Clifford gates (bottom). Top: $\rho_w = Z^{\otimes w}I^{\otimes n-w}$ represents $n$ distinct Pauli states, $\mathcal{C}_i$ is a 1-qubit Clifford operation in $\mathcal{C}_1^{\otimes n}$, and $\Pi_n$ is a permutation operation. Bottom: $\rho_{i} = \mathcal{C}_i \Pi_n \rho_{w} \Pi_n^{\dagger} \mathcal{C}_i^{\dagger}$ spreads over the entire Pauli group $\mathcal{P}_n$, $\tilde{\mathcal{U}_c} = \mathcal{U}_c \circ \Lambda$ is the noisy Clifford gate, and $M_{\rho_i, \mathcal{U}_c} = \mathcal{U}_c \rho_{i} \mathcal{U}_c^{\dagger}$. (b) Molecular structure of Dichloro-cyclobutanone, where C$_1$ to C$_7$ form a 7-qubit system. (c) Pulse sequence for the creation of labeled PPS via the method in Ref. \cite{Knill2000}. It consists of three parts: encoding, coherence selection (CS) and decoding. $\mathcal{U}_{c}$, realized by a 80 ms GRAPE pulse, is the Clifford gate to be certified. The instantaneous states are (unnormalized) $\rho_1 = I^{\otimes 6}\otimes Z$,  $\rho_2 = Z^{\otimes 7}$, $\rho_3 = \ket{0}\bra{0}^{\otimes 7}+\ket{1}\bra{1}^{\otimes 7}$, and $\rho_4 =  \ket{0}\bra{0}^{\otimes 6} \otimes Z_7$, respectively. (d) Experimental result for the certification of $\mathcal{U}_{c}$. $k=3^w\binom{7}{w}$ is the number of Pauli operators for weight $w$, while $k_w$ is the number of experiments via the sampling; $t$ is the typical time for the input Pauli state preparation, and $F_i$ is the calibration to capture the errors in preparation and measurement; $F_e$ is the experimental result of the probability of no error, and $F_{d,c}$ is the same quantity but without decoherence effect $E_d$. (e) Relationship among the experimental remaining signals (blue), decoherence effects (orange) and gate imperfections (gray) for different $w$.}}\label{everything}
\end{figure*}

To obtain $\text{Pr}(0)$, we can start from the input state $\ket{0}^{\otimes n}$, apply the $\mathcal{C}_1^{\otimes n}$ twirled channel, and measure the output state in the $n$-bit string basis \cite{Emerson2007}.  Equivalently, for an ensemble system we can replace $\ket{0}^{\otimes n}$ by $n$ distinct input states $\rho_w = Z^{\otimes w}I^{\otimes n-w}$ where $Z$ represents the Pauli matrix $\sigma_z$, followed by a permutation operation $\Pi_n$, and measure accordingly as shown in Fig. \ref{everything}(a).

From an experimental perspective this is still a difficult task. Ideally we want to make as few assumptions as possible about the ability to perform arbitrary Clifford operations since in practice  we can only implement $\tilde{\mathcal{U}_c} =\Lambda\circ \mathcal{U}_c$.  Moussa \emph{et al.}  \cite{Moussa2012} modified the  original twirling protocol in the following way.  By inserting the identity $\mathcal{U}_c \circ \mathcal{U}_c^\dagger$ appropriately, the circuit depicted in the upper panel of Fig. \ref{everything}(a) can be transformed to the lower one.   The input state $\rho_{i} = \mathcal{C}_i \circ \Pi_n( \rho_{w})$ is the input Pauli operator %spread over the entire Pauli group $\mathcal{P}_n$, and
 and the measurement $M_{\rho_i, \mathcal{U}_c} = \mathcal{U}_c( \rho_{i})$ (that can be calculated efficiently \cite{Aaronson2004}),
is also a Pauli operator.

By implementing the circuit in the lower panel of Fig. \ref{everything}(a), the probability of no error is \cite{Alex2013} %\R{[The factor of $1/2^n$ should be clarified]}
\begin{align} \label{noerror}
\text{Pr}(0) = \frac{1}{4^n} \left( 1+ \frac{1}{2^n}\sum_{i=1}^{4^n-1} \text{Tr}\left( \tilde{\mathcal{U}_c} \left( \rho_{i}\right) M_{\rho_i, \mathcal{U}_c}\right)\right).
\end{align}
Then substituting Eq. \eqref{noerror} to Eq. \eqref{fidelity_pr} will yield the average fidelity of the faulty Clifford gate $\tilde{\mathcal{U}_c}$.

Note that the above twirling protocol is limited to the certification of Clifford gates. For a general unitary gate, it is often impractical to realize the measurement operator $M_{\rho_i, \mathcal{U}} = \mathcal{U} \rho_{i} \mathcal{U}^{\dagger}$, whereas for a Clifford gate it can be decomposed  efficiently \cite{Aaronson2004}. It is possible to develop fault-tolerant  quantum computing where Clifford gates and  magic state preparation are the basic building blocks \cite{Bravyi2005,Souza2011}. In  these architectures   they are the only gates that need to be benchmarked \footnote{Note that benchmarking of measurements and preparation is unavoidable.}. For example, the encoding operation of the 3-qubit quantum error correction code is a Clifford gate comprising two controlled-NOT (CNOT) gates and a single qubit Hadamard gate, and has been certified in a 3-qubit solid-state NMR system \cite{Moussa2012}.

In spite of the simplification of the aforementioned way to estimate the average fidelity of Clifford gates, the complexity remains exponential as $4^n-1$ distinct Pauli states need to be prepared. Actually, measuring all of the expectation values is unnecessary if one only desires to approximate the average with a given confidence level and confidence interval \cite{Emerson2007}.  Hoeffding's inequality \cite{Venkatesh2012} states that if $x_1, . . . ,x_m$ are independent realizations of a random variable $x$, confined to the interval $[a, b]$ and with statistical mean $\mathbb{E}(x) = \mu$, then for any $\delta >0$ we have
\begin{align} \label{Hoeffding}
\text{Prob} \left( |\bar{x}-\mu| > \delta \right) \leq 2e^{-2\delta^2m/(b-a)^2},
\end{align}
where $\bar{x} = \frac{1}{m}\sum_{i=1}^m x_i$ is the estimator of the exact mean $\mu$, and $\text{Prob}(\epsilon)$ denotes the probability of event $\epsilon: |\bar{x}-\mu| > \delta$ which we want to minimize. Explicitly,  Hoeffding's inequality provides an upper bound on the probability that the estimated mean is off by a value greater than $\delta$. The confidence level and confidence interval are  $1-\text{Prob}(\epsilon)$ and [$-\delta, \delta$], respectively.

{When $\mu$ is the  average fidelity we have  $a=0$ and $b=1$.} Hence, for a given $\text{Prob}(\epsilon)$ and $\delta$, the number of experiments {calculated  by taking the $log$  of Eq. \eqref{Hoeffding} is}
\begin{align} \label{exp_number}
m\leq \frac{\text{ln}(2/\text{Prob}(\epsilon))}{2\delta^2}.
\end{align}
 Note that the number of experiments is independent of number of qubits  $n$, once the desired $\text{Prob}(\epsilon)$ and $\delta$ have been given. This result reveals that the estimation of the average fidelity of Clifford gates via twirling protocol is efficient and scalable. For instance, given a $99\%$ confidence level, \emph{i.e.,} $\text{Prob}(\epsilon)=1\%$ and $\delta = 0.04$, the total number of experiments is 1656, {independent of $n$}.

\paragraph*{Experiment.}
In the experiment we chose  $\mathcal{U}_{c}$ to be the Cifford gate used  to generate  maximal (7-qubit) coherence from single (1-qubit) coherence, up to single-qubit gates. It evolves $ZI^{\otimes n-1}$ to $Z^{\otimes n}$ and is the basic encoding process  for the pseudo-pure state (PPS) preparation method of Ref. \cite{Knill2000} shown in Fig. \ref{everything}(c). It  also plays a role  in  the creation of cat states. %$\left( \ket{0}^{\otimes n}+\ket{1}^{\otimes n} \right)/\sqrt{2}$ .
 The gate can be decomposed into a sequence of elementary Clifford gates of the type
\begin{align} \label{decomp}
e^{-i\frac{\pi}{4}X_i}e^{-i\frac{\pi}{4}Z_i Z_j} e^{-i\frac{\pi}{4}Y_i},
\end{align}
that  increase the order of coherence by evolving $Z_i$ to $Z_i Z_j$. Implementing $\mathcal{U}_{c}$  in experiment is nontrivial as it requires $2(n-1)$ single qubit operations and $(n-1)$ 2-qubit operations.

Our 7-qubit NMR processor is the per-$^{13}$C-labeled dichlorocyclobutanone derivative \cite{Johnson2008} shown in Fig. \ref{everything}(b) dissolved in
d6-acetone. The carbon nuclei labeled C$_1$ to C$_7$ denote the seven qubits.  Details of the molecular structure can be found in the Supplementary Material \cite{supplement}. $^1$H nuclei were decoupled by the Waltz-16 sequence throughout all experiments. The internal Hamiltonian of this system can be described as
\begin{align}\label{Hamiltonian}
\mathcal{H}_{int}=\sum\limits_{j=1}^7 {\pi \nu _j } Z_j  + \sum\limits_{j < k,=1}^7 {\frac{\pi}{2}} J_{jk} Z_j Z_k,
\end{align}
where $\nu_j$ is the resonance frequency of the \emph{j}th spin and $\emph{J}_{jk}$ is the scalar coupling strength between spins \emph{j} and \emph{k}. All experiments were conducted on a Bruker DRX 700 MHz spectrometer at room temperature.%See supplemental material for the values of all parameters.

The entire procedure to estimate the average fidelity of $\mathcal{U}_{c}$ can be divided into four parts, as follows:

(i) Sampling. To achieve a confidence level 99\% and precision $\delta = 0.04$, we computed that the required number of experiments is 1656 via Eq. \eqref{exp_number}. Then we randomly sampled  1656 distinct Pauli states out of the entire 7-qubit Pauli group, which has in total $4^7-1=16383$ elements. We distributed all 1656 input Pauli states to seven subgroups according to their Pauli weights $w=1$ to $w=7$. The primary reason for this distribution is that a quantum gate such as $\mathcal{U}_{c}$ here is usually more prone to error when applied to higher weight Pauli states. Additionally, the preparations of input Pauli states with different weights $w$ are distinct.

The sampling result is shown in Fig. \ref{everything}(d), where the number of sampled experiments $k_w$ for weight $w$ is around one tenth of the total number $k = 3^w\binom{n}{w}$.

(ii) Preparation and Calibration. For the creation of every input Pauli state, we employed an efficient sequence compiling program \cite{Ryan2008} to produce the corresponding pulse sequence. All pulses in the preparation sequences are selective and generated by Gaussian shapes. We then compared the state preparation results with the thermal equilibrium state as a calibration of the certification procedure, aiming to capture the errors in preparation and measurements. The typical duration $t$ for preparing a weight $w$ Pauli state and the related calibration results $F_i$ are both listed in Fig. \ref{everything}(d).

(iii) Evolution. The target operation $\mathcal{U}_{c}$ was optimized by a GRadient Ascent Pulse Engineering (GRAPE) pulse \cite{Khaneja2005}. Utilizing the GRAPE algorithm guarantees that $\mathcal{U}_{c}$ is a Clifford gate to a very good approximation, as traditional state-dependent shape pulses for multiple qubits in NMR are unlikely to form a strict Clifford operation. The GRAPE pulse of $\mathcal{U}_{c}$ was obtained with the pulse width chosen as 80 ms and a simulated fidelity of 0.99. A special calibration method was used in the experiment to ensure that the pulse acting on the spins was a very close approximation to the simulated (theoretical) pulse \cite{Weinstein2004}.

(iv) Measurement. After applying the GRAPE pulse of $\mathcal{U}_{c}$ to each input Pauli state in the experiment, we measured the corresponding output Pauli state by local readout pulses, and recorded the ratio of the remaining signal to that of the reference input state.  Next we averaged the results with respect to different weights $w$, as shown by $F_e$ in Fig. \ref{everything}(d). It is expected that the ratio will decrease as $w$ increases, since  higher coherences are less robust to the decoherence occurring during   $\mathcal{U}_{c}$.

\begin{figure}[htb]
\begin{center}
\includegraphics[width=\columnwidth]{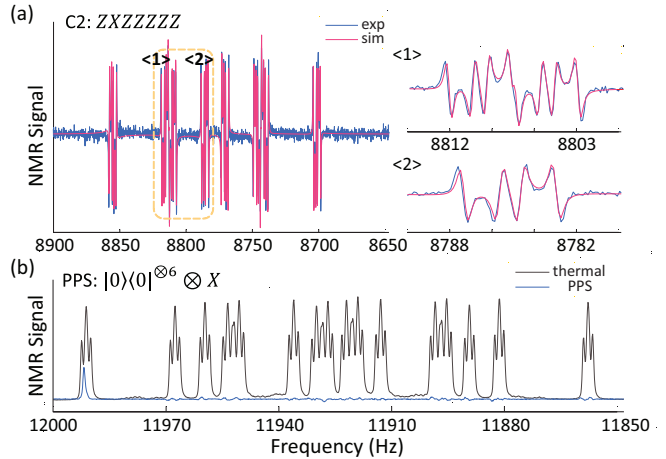}
\end{center}
\setlength{\abovecaptionskip}{-0.35cm}
\caption{\footnotesize{(color online). (a) NMR spectrum of $Z^{\otimes 7}$ under the observation of C$_2$. The simulated (red) spectrum is rescaled for lineshape comparison with the experimental (blue) one. (b) PPS spectrum (blue) based on the network in Fig. \ref{everything}(c), where $\mathcal{U}_{c}$ was employed as the encoding process. The spectrum of the thermal equilibrium state (black) is also shown.}}\label{spectra}
\end{figure}

 The probability of no error ($F_e$ in Fig. \ref{everything}(d)) is $\text{Pr}(0)\approx54.7\%$. The average fidelity of $\mathcal{U}_{c}$ via Eq. \eqref{fidelity_pr} is then $\bar{F}(\Lambda)\approx55.1\%$. {It is possible to decompose    $\mathcal{U}_{c}$ into  twelve 1-qubit gates and six 2-qubit gates \footnote{{The pulse was optimized using GRAPE, so the operation $\mathcal{U}_{c}$ is not really a sequence of 1- and 2-qubit  gates. However the initial guess for the pulse was generated from the 1-and 2-qubit gate sequence.}}}, so the average error per sub-gate is  $\simeq 2.5\%$. To quantify the decoherence contribution during $\mathcal{U}_{c}$, we followed the approach of phase damping \cite{Vandersypen2001} to simulate the dynamical process step by step. The average signal attenuation due to decoherence is shown by $E_d$ in Fig. \ref{everything}(d). Under the assumption that the decoherence error can be factorized,  the probability of no error $\text{Pr}(0)$ after theoretically removing the decoherence is 87.4\%, which means the average fidelity is 87.5\%. The average error  per sub-gate is  then $\simeq 0.7\%$. The remaining  errors are  mainly attributed to imperfection in the design and implementation of the GRAPE pulse.  Fig. \ref{everything}(e) shows  the relationship between the raw experimental results, decoherence effects and gate imperfections for each $w$.

 Fig. \ref{spectra}(a) shows the  spectrum of $Z^{\otimes 7}$  after  $\mathcal{U}_{c} (Z_7)$ under the observation of C$_2$. Comparing   the simulated and experimental spectra  gives a qualitative  indication of  the level of coherent control achieved in this 7-qubit system.  For another comparison we followed $\mathcal{U}_{c}$ by a set of  operations  to extract  the PPS as in Fig. \ref{everything}(c). The PPS spectrum by observing the labeled spin C$_7$ is shown in Fig. \ref{spectra}(b).

\paragraph*{Conclusion.}
We estimated  the average fidelity of a non-trivial 7-qubit Clifford gate using a twirling protocol together with a  random sampling method.  This is the largest gate-characterization reported in an experiment to date. The experimental spectra demonstrate reliable coherent control of this 7-qubit system while our benchmarking protocol gives an average gate fidelity of 55.1\% before accounting for decoherence, and 87.5\% after theoretically removing the contribution of decoherence. An important feature of the protocol is the relatively small number of experiments required: 1656 ($<2^{11}$) compared to $2.7 \times 10^8 (\approx2^{28})$ for process tomography. With further developments in experimental quantum information processing we expect that the methods used here will become standard tools for characterizing gate fidelities in larger processors.

We thank S. Y. Hou, O. Moussa, H. Park and G. R. Feng  for helpful comments and discussions. This work is supported by Industry Canada, NSERC and CIFAR. HL and GLL are supported by National Natural Science Foundation of China under Grant Nos. 11175094  and 91221205, the National Basic Research Program of China under Grant No. 2011CB9216002.


\begin{thebibliography}{99}
\bibitem{Preskill1998} J. Preskill, Proc. R. Soc. A \textbf{454}, 385 (1998).
\bibitem{Chuang1997} I. Chuang and M. Nielsen, J. Mod. Opt. \textbf{44}, 2455 (1997).
\bibitem{Poyatos1997} J. Poyatos, J. Cirac, and P. Zoller, Phys. Rev. Lett. \textbf{78}, 390 (1997).
\bibitem{Childs2001} A. Childs, I. Chuang, and D. Leung, Phys. Rev. A \textbf{64}, 012314 (2001).
\bibitem{Weinstein2004} Y. Weinstein, T. Havel, J. Emerson, N. Boulant, M. Saraceno, S. Lloyd, and D. Cory, J. Chem. Phys. \textbf{121}, 6117 (2004).
\bibitem{Brien2004} J. O'Brien, G. Pryde, A. Gilchrist, D. James, N. Langford, T. Ralph, and A. White, Phys. Rev. Lett. \textbf{93}, 080502 (2004).
\bibitem{Riebe2006} M. Riebe, K. Kim, P. Schindler, T. Monz, P. Schmidt, T. K\"{o}rber, W. H\"{a}nsel, H. H\"{a}ffner, C. Roos, and R. Blatt, Phys. Rev. Lett. \textbf{97}, 220407 (2006).
%\bibitem{Chow2009} J. Chow, J. Gambetta, L. Tornberg, J. Koch, L. Bishop, A. Houck, B. Johnson, L. Frunzio, S. Girvin, and R. Schoelkopf, Phys. Rev. Lett. \textbf{102}, 090502 (2009).
\bibitem{Chow2009} J. Chow \emph{et al.}, Phys. Rev. Lett. \textbf{102}, 090502 (2009).
%\bibitem{Bialczak2010} R. Bialczak, M. Ansmann,	 M. Hofheinz,	E. Lucero, M. Neeley,	 A. O'Connell, D. Sank, H. Wang, J. Wenner,	M. Steffen,	A. Cleland, and J. Martinis, Nat. Phys. \textbf{6}, 409 (2010).
\bibitem{Bialczak2010} R. Bialczak \emph{et al.}, Nat. Phys. \textbf{6}, 409 (2010).
\bibitem{Kim2014} D. Kim \emph{et al.}, Nature \textbf{511}, 70 (2014).
\bibitem{Feng2013} G. Feng, G.  Xu, and G. Long,  Phys. Rev. Lett. \textbf{110}, 190501 (2013).
\bibitem{Emerson2005} J. Emerson, R. Alicki, and K. Zyczkowski, J. Opt. B \textbf{7}, S347 (2005).
\bibitem{Knill2008} E. Knill \emph{et al.}, Phys. Rev. A \textbf{77}, 012307 (2008).
\bibitem{Ryan2009} C. Ryan, M. Laforest, and R. Laflamme, New J. Phys. \textbf{11}, 013034 (2009).
\bibitem{Emerson2007} J. Emerson \emph{et al.}, Science \textbf{317}, 1893 (2007).
\bibitem{Dankert2009} C. Dankert, R. Cleve, J. Emerson, and E. Livine, Phys. Rev. A \textbf{80}, 012304 (2009).
\bibitem{Moussa2012} O. Moussa, M. Silva, C. Ryan, and R. Laflamme, Phys. Rev. Lett. \textbf{109}, 070504 (2012).
\bibitem{Flammia2011} S. Flammia and Y. Liu, Phys. Rev. Lett. \textbf{106}, 230501 (2011).
\bibitem{Silva2011} M. Silva, O. Landon-Cardinal, and D. Poulin, Phys. Rev. Lett. \textbf{107}, 210404 (2011).
\bibitem{Bennett1996} C. Bennett, D. DiVincenzo, J. Smolin, W. Wootters, Phys. Rev. A \textbf{54}, 3824 (1996).
\bibitem{Silva2008} M. Silva, PhD thesis, University of Waterloo, 2008.
\bibitem{Aaronson2004} S. Aaronson and D. Gottesman, Phys. Rev. A \textbf{70}, 052328 (2004).
\bibitem{Bravyi2005} S. Bravyi and A. Kitaev, Phys. Rev. A \textbf{71}, 022316 (2005).
\bibitem{Souza2011} A. Souza, J. Zhang, C. Ryan, and R. Laflamme, Nat. Comm. \textbf{2}, 169 (2011).
\bibitem{Alex2013} D. Trottier, Master thesis, University of Waterloo, 2013.
\bibitem{Venkatesh2012} S. Venkatesh, \emph{The Theory of Probability: Explorations and Applications}. Cambridge University Press, 2012.
\bibitem{Knill2000} E. Knill, R. Laflamme, R. Martinez, C. Tseng, Nature \textbf{404}, 368 (2000).
\bibitem{Johnson2008} J. W. Johnson, D. P. Evanoff, M. E. Savard, G. Lange, T. R.  Ramadhar, A. Assoud, N. J. Taylor, and G. I. Dmitrienko, J. Org. Chem. \textbf{73}, 6970 (2008).
\bibitem{supplement} See supplementary material for more information.
\bibitem{Ryan2008} C. Ryan \emph{et al.}, Phys. Rev. A \textbf{78}, 012328 (2008).
\bibitem{Khaneja2005} N. Khaneja \emph{et al.}, J. Magn. Reson. \textbf{172}, 296 (2005).
\bibitem{Vandersypen2001} L. Vandersypen \emph{et al.}, Nature \textbf{414}, 883 (2001).
\end{thebibliography}
\end{document}